\documentclass[a4paper,fleqn,usenatbib]{mnras}
\usepackage[T1]{fontenc}
\usepackage{ae,aecompl}
\usepackage{multirow,ulem}

\usepackage{graphicx}	% Including figure files
\usepackage{amsmath}	% Advanced maths commands
\usepackage{amssymb}	% Extra maths symbols
\usepackage{natbib}

\usepackage[dvipsnames]{color}
\definecolor{webgreen}{rgb}{0,.5,0}
\newcommand{\ufhref}[3][blue]{\href{#2}{\color{#1}{#3}}}%

\title[Diffuse Galactic Gamma-rays from star clusters]{Diffuse Galactic Gamma-rays from star clusters}
\author[Nath, Eichler]
{
Biman B. Nath$^{1}$, David Eichler$^{2}$ \\
\footnotesize \it $^{1}$Raman Research Institute, Sadashiva Nagar, Bangalore 560080, India \\ 
\footnotesize \it $^{2}$Department of Physics, Ben-Gurion University, Beer Sheva, Israel \\
}
\date{Accepted XXX. Received YYY; in original form ZZZ}

\pubyear{2020}

\begin{document}
\label{firstpage}
\pagerange{\pageref{firstpage}--\pageref{lastpage}}
\maketitle

%%%%%%%%%%%%%%%%%%%%%%%%%%%%%%%%%%%%%%%%%%%%%%%%%%%%%%%%%%%%%%%%%%%%%%%%%

\newcommand{\3}{\ss}
\newcommand{\n}{\noindent}
\newcommand{\eps}{\varepsilon}
\def\be{\begin{equation}}
\def\ee{\end{equation}}
\def\ba{\begin{eqnarray}}
\def\ea{\end{eqnarray}}
\def\de{\partial}
\def\msun{M_\odot}
\def\div{\nabla\cdot}
\def\grad{\nabla}
\def\rot{\nabla\times}
\def\ltsima{$\; \buildrel < \over \sim \;$}
\def\simlt{\lower.5ex\hbox{\ltsima}}
\def\gtsima{$\; \buildrel > \over \sim \;$}
\def\simgt{\lower.5ex\hbox{\gtsima}}
\def\etal{{et al.\ }}

%%%%%%%%%%%%%%%%%%%%%%%%%%%%%%%%%%%%%%%%%%%%%%%%%%%%%%%%%%%%%%%%%%%%%%%%%

%\title{Galactic diffuse gamma-ray emission from star clusters}

%\author{Biman B. Nath}
%\affil{Raman Research Institute, Sadashiva Nagar, Bangalore, India}
%\email{biman@rri.res.in}

%\author{Nath, Eichler}

\begin{abstract}
We demonstrate that young star clusters  have  a $\gamma$-ray surface  brightness comparable to that of the diffuse Galactic emission (DGE), and estimate that their sky coverage in the direction of the inner Galaxy  exceeds unity. We therefore suggest that they comprise a significant fraction of the DGE. 
\end{abstract}


\section{Introduction}
The diffuse Galactic $\gamma$-ray emission (DGE) has been a test bed for the theories of cosmic ray (CR) acceleration sites and transport, ever
since \cite{morrison1958} pointed out the connection between them. The standard paradigm of CR acceleration site has been that supernovae (SN)
remnant (SNR) accelerate CR particles from the interstellar medium (ISM) material in Sedov-Taylor phase. These CR particles 
interact with protons in the ISM to produce pions and then neutral pions decay to produce $\gamma$-rays. Therefore CR grammage is tracked by $\gamma$-ray production in the ISM. In the standard scenario of CR acceleration in SNRs, the grammage  traversed by  CRs is accumulated as they propagate through the ISM.

After the initial observations of {\it OSO-3}, {\it SAS-II}, 
studies with {\it COS-B} and thereafter {\it EGRET} aboard the Compton Gamma-Ray Observatory
helped improve the knowledge of DGE in GeV range. Analysis of data \citep{bloemen1989, hunter1997} showed that pion decay through CR interaction with ISM protons is the major contributor
to the emission in the GeV range, while bremsstrahlung and inverse -Compton scattering may be important in MeV range. 
While CR particles from SNRs are thought to explain the DGE in general, this paradigm is not without its share of problems, especially when the DGE in the inner Galactic region is considered. The predicted intensity falls short of the observed values. It is possible that Galactic CR spectrum is different from that observed in the solar neighbourhood, because a harder spectrum is required, or that there exists an unresolved population of SNRs \citep{hunter1997b, weekes1997}.  \citet{berezhko2000} estimated that such unresolved SNRs would contribute $\le 10\%$ to the total DGE a $\sim 1$ GeV, but this fraction would be larger at higher energies.

With regard to both these points, of DGE requiring additional sources, with harder spectrum, the
%However, 
recent detections of young star clusters in $\gamma$-rays \citep{Ackermann2011, Aharonian2019} raises the possibility that they can be 
important sources. Although they have been considered in some analysis \citep{ackermann2012} of DGE, they have been treated as sources of CR that
would interact with ISM protons. Recently, \cite{gupta2018} have shown that the wind termination shocks (WTS) of young, massive compact star clusters can accelerate CRs and  explain the $\gamma$-ray, X-ray  and radio luminosities of star clusters, as observed. They calculate a conversion factor between the mechanical luminosity of winds and $\gamma$-ray is roughly $L_\gamma/L_w \approx 10^{-3}\hbox{--}10^{-2}$, which is consistent with observations \citep{Ackermann2011}.  This calculation is reviewed below.   \citep{gupta2020}
 also show that the expected combination of WTS  and  SNe shocks in star clusters, given  the conventional wisdom about stellar evolution and stellar initial mass functions, can explain the Neon isotope ratio. More importantly, they have shown that CRs can obtain most of their grammage within the superbubble of the star clusters, mainly in the shocked-wind and shocked ISM region.  This implies that most of the $\gamma$-rays should be generated locally, within the superbubble.  

A further indication that star clusters may contribute  a substantial fraction of DGE, comes from the analysis by \cite{deboer2017}, who
found that molecular clouds can explain the DGE flux towards the central molecular zone, with a harder spectrum than that found elsewhere. Interestingly, the observations of \cite{Aharonian2019} reveal that $\gamma$-ray spectrum towards star clusters have a harder spectrum.

The energy budget of CRs (and consequently that of $\gamma$-rays produced by them) can be estimated as follows. Extrapolating from a census of  over 400 O3-B2 stars in the solar neighbourhood, the number of OB stars within the solar circle has been estimated to be $\sim 2\times 10^5$ \citep{Reed2005}, and the average mechanical power associated with stellar winds arising from them is $L_w \approx 10^{36}$ erg s$^{-1}$. Therefore the total mechanical power associated with WTS is $\approx 2 \times 10^{41}$ erg s$^{-1}$ (see also \citet{kang2018}), compared to $\approx 6 \times 10^{41}$ erg s$^{-1}$ of SNe shocks, corresponding to a SN rate of $\sim 2$ per century \citep{Diehl2006}.  
The fraction of $L_w$ that is radiated in $\gamma$-rays, via CR particles accelerated in WTS and their interaction with the dense shell material near the outer shock, depends on many factors, including the ambient density (see below for details), and observationally it can range between $0.002\hbox{--}0.01$ (see examples below), except in the case of Cygnus and Westerlund 1 where it is $\le 3 \times 10^{-4}$. This implies a minimum $\gamma$-ray luminosity from OB stars of $\ge 4 \times 10^{38}$ erg s$^{-1}$. This is a substantial fraction of the total $\gamma$-ray luminosity of $\approx 10^{39}$ erg s$^{-1}$ of our Galaxy, and tantalizingly close to the amount required to fill the gap between the observed and prediction from leaky-box model.

Here we consider the possibility that much of the $\gamma$-ray emission from the Galaxy is from superbubbles and whether this emission should appear as separate sources or as diffuse (if not completely smooth) emission.
After reviewing the energetics, %more carefully, 
we  show that the sky filling factor of star clusters in the inner Galactic region is large. Then we show that the size of the star clusters scale as $r_s \propto L_w^{1/2}\propto L_\gamma^{1/2}$ (which follows from the evolution of stellar wind inflated bubbles in ISM). In other words, every line of sight will not only intercept a young ($\le 10$ Myr) star cluster, it will also have the same specific intensity, which turns out to be in the ballpark of the observed DGE. The only parameters involved here are superbubble evolution (based on \citet{weaver1977} model) and the
observed scaling of $\gamma$-ray luminosity with mechanical luminosity of star clusters. We also estimate the anisotropy of DGE expected in this scenario.

\section{Why Do We Care?}
Why do we care whether the diffuse Galactic $\gamma$-ray emission is dominated by localized, overlapping sources as opposed to a smooth distribution of nuclear collisions in the Galactic disk at large?  We care because the DGE is a clean diagnostic of cosmic ray propagation in the Galaxy.  The traditional leaky box model, by definition, assumes that the escape time  from the Galactic volume is longer than the time needed for the cosmic rays from a given source to homogenize their distribution in the Galactic disk.       However this assumption may be false. It may be that cosmic rays escape the Galactic disk  not far from their source at characteristic distance $d$ from their source. In this case, the distribution of cosmic rays in the Galactic and hence the collisions they undergo   occur mostly with $\lesssim d$ of their source,   and the $\gamma$-rays emitted in those  collisions appear as a localized (but not necessarily point) source rather than getting smeared across the entire Galactic disk. Now if those   localized sources overlap, the overall emission may  appear mostly diffuse. On the other hand, even if the cosmic rays from a given source filled the disk homogeneously, the $\gamma$-ray emission from their collisions could appear splotchy if the baryon distribution in the disk is splotchy. So using $\gamma$-ray emission to distinguish between different scenarios for cosmic ray distribution requires care.

The leaky box has in recent year encountered problems accounting for the flat antiproton to proton ratio in the  several hundred GeV range. This is because the secondary boron to primary carbon ratio in the Galactic disk has been observed to decrease with energy out to $10^2$ GeV, and has been traditionally attributed to an escape rate from the Galaxy that increases with energy. The apparently paradoxical absence of a decrease of the  antiproton to proton  ratio, would, in the absence of new physics, beg for an alternative explanation because antiprotons are also presumed to be secondaries. \citet{eichler2017} suggested that some of the decrease in the secondary to primary ratio is due to a correlation between grammage traversed and maximum acceleration energy, leading to a decrease in the secondary boron to primary carbon ratio if the cosmic rays are due to an assortment of sources with varying maximum energies. It was further noted in \citet{eichler2017} that the maximum  energy attainable by shock acceleration, if established by Alfv\'en wave damping upstream of the shock by ion-neutral collisions, must be less than $\sim 10^{12}$ eV; above this energy, this damping mechanism is unlikely to operate. So if this is the reason for the secondary to primary ratio decrease below 100 GeV per nucleon, then flattening of the ratio above $100$ GeV per nucleon is expected {\it a priori}, and is confirmed by observations. The de-coupling of grammage and CR propagation in the ISM, as opposed to the grammage accumulated within source regions, has also been advocated in this regard by \cite{cowsik2016}. As the review by \citet{gabici2019} have emphasized, the de-coupling of grammage with CR propagation in ISM would solve many outstanding problems in CR phenomenology.

%\section{Preliminaries}
\citet{gupta2020} have noted that the relative contributions to the Galactic gamma ray luminosity from WTS and from supernovae can be estimated a priori from the paradigm of stellar evolution and initial mass spectra. %They assume that 
The wind power from an OB star is $10^{36}$ erg/s over a life span of $(3 \hbox{--} 10)$ Myr
%{Newcut}\times 10^6$ years 
giving a total energy output of $(1\hbox{--}3) \times 10^{50}$ erg compared to $10^{51}$ ergs from a supernova. On the other hand, the supernova shock  expands much more quickly so that the cosmic rays it produces suffer adiabatic losses, whereas the termination shocks maintain a steady stand-off in pressure equilibrium with the interstellar medium,  and the cosmic rays escape from the WTS through nearly stationary material and do not suffer adiabatic losses, so one might expect the contributions to the cosmic rays and attendant gamma rays to be comparable.  This has been confirmed by detailed numerical computation including the fusion of many winds and supernovae within clusters \citep{gupta2020}.

%A detailed numerical computation, including the fusion of many winds and supernovae within star forming clusters was performed by \citet{gupta2020} and this is confirmed. 

   In the present paper, we are concerned with the total area $4\pi R_o^2$ subtended by the  outer shock region where $\gamma$-rays are produced. 
Consider the WTS with a mass loss rate 
$\dot M = 4\pi \rho(R)R_{wts}^2u$.
The ram pressure $\rho(R)u(R)^2= \dot M u/(4\pi R_{wts}^2)$ at this radius equals the pressure in the shocked wind region $P_{in}$, so the area of the WTS is
$\pi R_{wts}^2 = \dot M u/(4P_{in})=L_w/(2 u P_{in})$, where $L_w=0.5 \dot M u^2$ is the wind mechanical power. The forward shock distance is larger 
than the WTS by a factor $R_o/R_{wts}\approx  29 (L_w/10^{36} \, {\rm erg/s})^{-1/10} t_{Myr}^{1/5} (n_{ISM}/cc))^{1/10}$ \citep{sharma2014}, and the 
relevant area is therefore $\pi R_o^2=(L_w/(2 u P_{in}))(R_o/R_{wts})^2$.
Now consider the sum of any number of winds that may merge before reaching a collective termination shock.  If both mass and energy are conserved in 
the mergers (i.e. no energy lost to radiation), then the total mass loss rate $M_{tot}$ is the sum of the individual ones $\dot{M_i}$, and the total power 
$\dot{M_{tot}} u_{final}^2$ is the sum of $\dot M_i u_i^2$, so $u_{final}$ is just the RMS of $u_i^2$ averaged over $\dot M$, which is at least the mean value of $u$.
%{\bf HERE IT WOULD BE GOOD TO TABULATE THE CORRECT VALUES IF THERE ARE ANY CATALOGUES OF OB STARS}. 
Taking $\dot{M}$ to be $10^{-6}$ solar masses per year, u to be $2 \times 10^8$ cm s$^{-1}$, and  $P_{in}\approx 10^{-10}$ dynes/cm$^2$ 
(as observed in superbubbles, \citet{lopez2014}),  and a total number of OB stars in the Galaxy of $2 \times 10^5$, 
we obtain an estimate of the total area of  $4.2 \times 10^{46}$  cm$^2$ (radiation loss would make this figure $\sim 1.5 \times 10^{46}$ cm$^2$, see below).

Now consider  the Galactic disk, which for a rough estimate can be taken as $R_{G} = 10R_{10}$ kpc   in radius and 100 pc in thickness.   The total area of its periphery is $2 \pi R_G^2 + 2\pi R_G h \simeq 5.6 \cdot 10^{45}R_{10}^2$ cm$^2$. 
So if the WTSs of OB stars were pinned to the furthest reach of the Galaxy along every line of sight, the fraction of sky it would occupy  would be more than unity by the above calculation. Moreover, if we assume the OB star is equally likely to be anywhere along the line of sight between us and the furthest reach, then  the fraction is much higher because the sky fraction covered by an object of distance d is proportional to $d^{-2}$. As this diverges at small d, we must introduce a cutoff $d_{min}$, that is of order the size of the WTS itself, some $30$ pc for $L_w\approx 10^{39}$ erg s$^{-1}$. The average enhancement of the sky coverage, relative to placing the source at the furthest reach, $d_{max}$, along the line of sight,  is  $d_{max}/d_{min}\sim 30$. So the solid angle subtended by the WTSs may be several skies rather than a fraction of a sky.

%\sout{\blue{
%Note further that the gamma rays are released from a larger sky area than subtended by the WTS. So while the diffuse Galactic emission may be non-uniform at some level, it is hard to quantitatively predict the level of nonuniformity. {\bf BUT BIMAN, SHOULD YOUR SIMULATIONS PROVIDE SOME ESTIMATE OF THE AREA PER UNIT MASS LOSS, AS THEY ARE PROPORTIONAL TO EACH OTHER, HENCE THEIR RATIO SHOULD BE ROBUST?}
%}\red{This is not needed anymore-- the outer shock and WTS estimates given above are verified in simulations.}}

\section{Superbubble size distribution}
Superbubbles in galaxies are believed to be produced by star clusters, whose luminosity function is observed to be a power-law \citep{mckee1997},
$
\phi(L) \propto L^{-2} dL  \,.
$
It has been found by \cite{oey1997} that the size distribution of HI holes, which are produced by superbubbles, can be explained in 
terms of  bubbles that are stalled by the pressure of the ambient interstellar medium. %While they estimated this size distribution in the 
%case of adiabatic expansion of bubbles, a 
A similar distribution also follows %has been shown to result 
even in the case of superbubbles dominated by radiative cooling \citep{nath2020}. The main result that follows from stalling of bubbles is that the radius of bubbles scales with luminosity as $r_s \propto L_w^{1/2}$, where $L_w ( \propto L$) denotes the mechanical power of the stellar wind.
as shown below.

The evolution of bubbles in the case of cooling can be written in terms of a parameter $\eta$ as follows,
\be
r_s (t) =({\eta L_w t^3\over \rho}) ^{1/5} \,.
\ee
%Here, $L_w$ is the mechanical power of the stellar wind and supernovae inside the cluster.
Hydrodynamical simulations indicate %show that the  value of the parameter $\eta$ can range between 
$\eta \sim 0.1\hbox{--}0.5$ \citep{sharma2014}. Observationally, it is found that the shell radius is roughly a fraction $\sim 0.6$ of the adiabatic case \cite{krause2014}, which indicates $\eta \sim 0.1$, a value we adopt here. %We adopt $\eta\approx 0.1$ here. 
The bubble stalls when the expansion speed becomes comparable to the ambient sound speed.
The stalling radius  (using adiabatic sound speed with $\gamma=5/3$) is,
\ba
r_s &\approx& 151 \, {\rm pc} \Bigl ({P_{\rm amb} \over 3.5 \times 10^{-12} \, {\rm dyne} \, {\rm cm}^{-2}} \Bigr )^{-3/4} \nonumber\\ && \Bigl ({L_w \over 10^{38} \, {\rm erg} \, {\rm s}^{-1}} \Bigr )^{1/2} \Bigl ({\rho \over 10^{-24} \, {\rm g} \, {\rm cm}^{-3}} \Bigr )^{1/4} \,,
\label{rs-lw}
\ea
where we have used an average pressure corresponding to %$3.5 \times 10^{-12}$ dyn cm$^{-2}$, or 
$P/k\approx 2.5 \times 10^4$ cm$^{-3}$ K \citep{jenkins2011}.
For the same parameters, the stalling time is $t_{\rm stall} \approx 1.6 \, (L_w/10^{38} \, {\rm erg} \, {\rm s}^{-1})^{1/2}$ Myr.
%{Newcut}The stalling time scale is given by,
%\ba
%t_{\rm stall}&\approx& 1.6 \, {\rm Myr} \Bigl ({P_{\rm amb} \over 3 \times 10^{-12} \, {\rm dyne} \, {\rm cm}^{-2}} \Bigr )^{-5/4} \nonumber\\ && \Bigl ({L_w \over 10^{38} \, {\rm erg} \, {\rm s}^{-1}} \Bigr )^{1/2} \Bigl ({\rho \over 10^{-24} \, {\rm g} \, {\rm cm}^{-3}} \Bigr )^{3/4} \,.
%\ea
Therefore, we find that the superbubbles stall within $\ll 10$ Myr even for the largest clusters, with $L_w \le 3\times 10^{39}$ erg s$^{-1}$ (corresponding to the number of OB stars $N_{\rm OB}\sim 3000$). In other words, clusters reach a stalling radius within the time scale that they are bright in $\gamma$-rays.

The relation between $r_s$ and $L_w$ leads to a size distribution. % in the following way. 
Suppose the superbubbles are produced at a constant rate (for
a constant star formation rate). The number of bubbles after a time $t$ with radii in the range $r_s$ to $r_s+dr_s$ will depend on $\phi(L_w)dL_w$.
%{Newcut}where $L_w$ and $dL_w$ are related to the radii in the above described manner. 
Therefore the differential size distribution will be given by
\citep{oey1997},
\be
N(r_s) \propto  \phi(L) \Bigl ( {\partial r_s \over \partial L_w} \Bigr )^{-1} \, \propto L_w^{-2+1/2} \propto r_s^{-3}\,,
\ee
which is the distribution we have used in Section 1. This size distribution has also been observationally confirmed by the {\it HI Nearby Galaxy Survey} %of $20$ nearby galaxies 
\citep{bagetakos2011}.

\section{Sky coverage for superbubbles}
%We first discuss the total energy budget of Galactic CRs and the contribution However, this assumption may be false from SNRs and star clusters. As shown by \citep{gupta2020, seo2018}, more than a quarter of the total CR luminosity likely arises from WTSs in young star clusters. Extrapolating from a census of  over 400 O3-B2 stars in the solar neighbourhood, the number of OB stars within the solar circle has been estimated to be $\sim 2\times 10^5$ \citep{Reed2005}, and the corresponding SN rate of $1\hbox{--}2$ per century. This is also consistent with the estimate of $1.9\pm1.1$ per century from $^{26}Al$ observations
%\citep{Diehl2006}. For an average mechanical power of stellar wind per OB star of $\sim 10^{36}$ erg s$^{-1}$, the total power of WTSs is given by
%$2\times 10^{41}$ erg s$^{-1}$. Using the fact that a fraction $(1\hbox{--}5)\times 10^{-3}$ of the mechanical power is converted in GeV $\gamma$-rays,
%the total $\gamma$-ray luminosity of the Galaxy is then given by $(2\hbox{--}10) \times 10^{38}$ erg s$^{-1}$. %, using the lower value). 
%If this constitutes a quarter of the total energy budget, then the rest would amount to $\sim 6 \times 10^{41}$ erg s$^{-1}$, corresponding to the above mentioned SN rate, with $10^{51}$ erg per SN. 
%Therefore WTSs can supply a significant fraction of the total Galactic $\gamma$-ray luminosity of $10^{39}$ erg s$^{-1}$ \citep{strong2010}. 
We now estimate the sky coverage in detail, taking into account their spatial distribution in the Galaxy.
According to \citet{gupta2020}, WTS in massive, compact star clusters can produce significant $\gamma$-rays. They found that clusters with $N_{\rm OB}\ge 10$ and that are younger than $\approx 10$ Myr  are especially suitable  for the Mach number of WTS to be large enough and consequently for  $\gamma$-ray emission. %for the production of $\gamma$-rays at individually detectable levels.

%We can estimate the number of star clusters from this census of OB stars in the inner Galactic region, mentioned above, using the cluster luminosity function, 
%$dN  =A N_{\rm OB}^{-2} dN_{\rm OB}$\citep{mckee1997}, where $N_{\rm OB}$ is the number of OB stars in a cluster, and $A$ is a normalising constant. If we take $10$ as the minimum number of OB stars,
%then using the fact that the largest OB association in Galaxy has $\sim 7000$ OB stars \citep{mckee1997},  the total number of OB stars is given by $
%A \ln (7000/10)$ which when equated to $2 \times 10^5$, gives the normalisation $A \sim 30000 $
%as the total number of clusters.

Next consider the spatial distribution of OB associations in the Milky Way. \cite{Bronfman2000} observed $748$ OB associations across the Galactic disk and found their distribution to peak at $R_p=0.75 R_0$,  $R_0=8.5$ kpc being the solar distance from the Galactic centre. We find that their inferred (differential) distribution can be roughly fitted by 
\be
dN(R) = \Sigma_0 e^{-{(R -R_p)^2\over \sigma ^2}} \, 2\pi R dR,
\label{ob_dist}
\ee
where $R$ is the galacto-centric distance and $\sigma\approx 3$ kpc. This fit yields a difference of a factor $\sim 4$ between the peak and the value at solar distance, as found by \cite{Bronfman2000} (dotted line in the bottom left panel of their Figure 3). Here, $\Sigma_0$ is a normalising factor, with dimensions of kpc$^{-2}$. 
The total number of clusters within the solar radius can be found by integrating this distribution. Numerically,
\be
\int_0 ^{R_0} R dR e^{{-{(R -R_p)^2\over \sigma ^2}}}\approx 26 \, {\rm kpc}^2 \,,
\ee
for $R_p=6.3$ kpc, $\sigma=3$ kpc, and $R_0=8.5$ kpc. Therefore the total number of clusters is $\approx 52\pi\Sigma_0$. This allows us to estimate $\Sigma_0$ 
in the following manner.

%%%%%%%%%%%%%%%%%%%%%%%%%%%%%%%%%%%%%%%%%%%%%%%%%%%%%%%%%%%
\begin{figure}
\vspace{24pt}
\includegraphics[width=80mm]{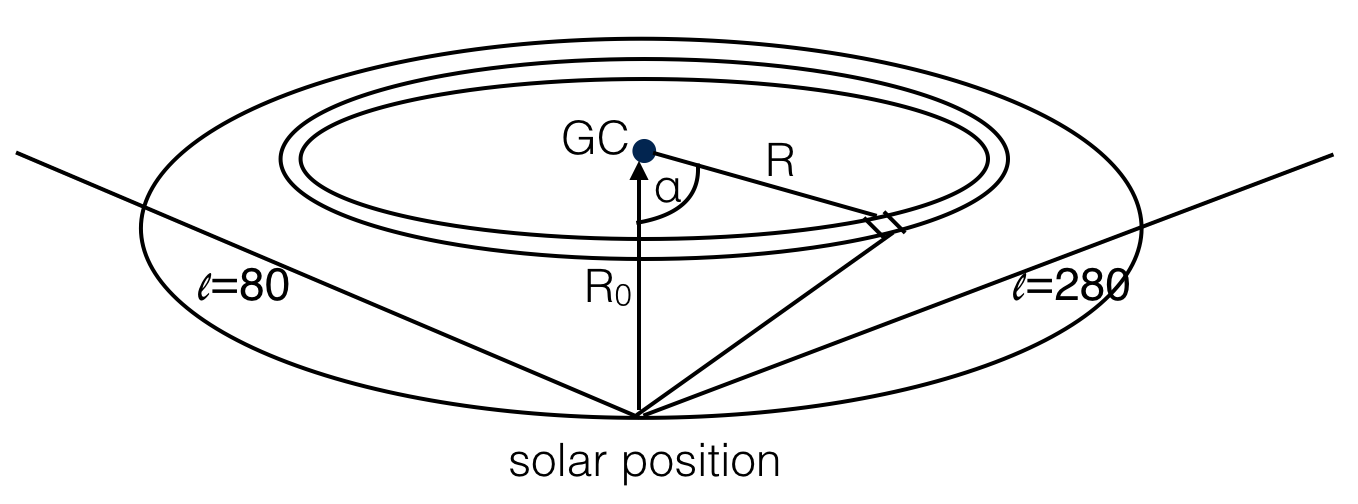}  %%%{specz25.eps}
\caption{Schematic diagram for calculation of sky coverage. GC denotes the Galactic centre, $R$ is the Galactocentric distance, and  the solar position is indicated. Also shown are the limiting longitudes ($80,280$) within which the inner Galactic region is viewed.
         }
\label{schem}
\end{figure}
%%%%%%%%%%%%%%%%%%%%%%%%%%%%%%%%%%%%%%%%%%%%%%%%%%%%%%%%%%%

We can take the Milky Way star formation rate (SFR) of $\sim 2$ M$_\odot$ yr$^{-1}$ \citep{chomiuk2011} for the last $\sim 10$ Myr. %\footnote{Note that these estimates of SFR are averaged over the lifetime of massive stars, $\sim 3\hbox{--}10$ Myr.} % Therefore this is a reasonable estimate.) 
This gives a total census of $2 \times 10^5$ OB stars,
for $1$ OB star per $\sim 100$ M$_\odot$ stellar mass using Kroupa Initial mass function. This is consistent with $2 \times 10^5$ OB stars found within solar circle
\citep{Reed2005}. The luminosity function of clusters is as mentioned earlier, with $L\propto N_{\rm OB}$. %given by $dN\propto N_{\rm OB}^{-2} dN_{\rm OB}$ \citep{mckee1997}. 
We adopt a minimum of $10$ OB stars for the production of $\gamma$-rays, while the largest OB association in our Galaxy has $N_{\rm OB} \sim 7000$ \citep{mckee1997}. This gives an average of 
 $N_{\rm OB} \approx 65$  per cluster. Together with the above estimate, we then have $\sim 3076$ OB associations younger than $\le 10$ Myr. 
This gives $\Sigma_0\approx 19$ kpc$^{-2}$.
These clusters will have a superbubble triggered by the winds of member OB stars, and the superbubbles will have a size distribution depending on how large the cluster is and on its age. As shown in the previous section, %We will show in Section 
the differential size distribution of superbubbles is given by $dN (r_s) \propto r_s^{-3}$. 
Together with the above described spatial distribution of clusters, we can write the 
combined size and spatial
distribution of superbubbles at Galactocentric radius $R$ as
\be
dN_s(r_s,R)  =C(R) r_s^{-3} \,dr_s\,
\ee
Here, the normalising constant $C(R)$ depends on $N(R)$, the number of clusters in an annulus of width $dR$ at Galactocentric radius $R$,
\be
 C(R)= N(R) \Bigl [ {1 \over r_{s,min}^2} - {1 \over r_{s,max}^2} \Bigr ]^{-1} \,,
\label{norm2}
\ee
 where 
$r_{s,min}, r_{s,max}$ are the minimum and maximum sizes. % of superbubbles. 

Let us estimate the sky filling factor of these clusters in the inner Galactic region, from the vantage point of the solar system. 
 Consider an annulus in the disk of width $dR$ at a Galacto-centric radius of $R$, and consider a differential area
element in that annulus at an angle $\alpha$ as defined in Figure \ref{schem} and of angular width $d\alpha$. The angle $\alpha$ is defined
with respect to the line connecting the Galactic centre and the solar location. Consider a superbubble of radius
$r_s$ there. The solid angle subtended by this superbubble at the solar location is given by, 
\be
\Omega (\alpha, R, r_s)= {\pi r_s^2 \over R_0^2 + R^2 \mp 2 R_0 R \cos \alpha}
\ee
where the minus sign is for $\alpha < \pi/2$ and plus for $\alpha > \pi /2$. Given the size distribution of superbubbles as mentioned above,
the total solid angle subtended by them from this area element is,
\ba
\Omega (\alpha, R) &=&\int C(R) r_s^{-3} {\pi r_s^2 \over R_0^2 + R^2 \mp 2 R_0 R \cos \alpha}dr_s\nonumber\\
&=&{N(R) dR d\alpha \over \Bigl [ r_{s,min}^{-2} -  r_{s,max}^{-2} \Bigr ]}\int{\pi r_s^{-1} \over R_0^2 + R^2 \mp 2 R_0 R \cos \alpha}dr_s \nonumber\\
&=& {\pi \Sigma_0 \ln [{r_{s,max} \over r_{s,min}}]  \over \Bigl [r_{s,min}^{-2} -  r_{s,max}^{-2} \Bigr ]}{ R e^{{-{(R -R_p)^2\over \sigma ^2}}}
dR d\alpha\over R_0^2 + R^2 \mp 2 R_0 R \cos \alpha} 
\ea
The total solid angle subtended by all clusters can be determined by integrating these over $R$ and $\alpha$. The limits on $\alpha$ depend on the relevant range of Galactic longitudes. Consider a range of Galactic longitudes
($l, -l$), then there would be annuli at large $R$ (nearer to $R_0$) for which the $\alpha$ integration would 
not be over $2\pi$, but over a limited section. The limits on the $\alpha$ integration are as follows:
\ba
&&\alpha =0\rightarrow 2 \pi \,,  \, R< R_0 \sin l \,, \nonumber\\
&& \alpha= {\pi \over 2} -l -\cos ^{-1} \Bigl ({R_0 \sin l \over R} \Bigr ) \rightarrow
-{\pi \over 2} +l + \cos ^{-1} \Bigl ({R_0 \sin l \over R} \Bigr )\,, \nonumber\\
&&\qquad\qquad  R> R_0 \sin l
\ea
Numerically integrating over $\alpha$ within these limits for $l=80^\circ$ (as used in the inner Galactic region by
 \cite{ackermann2012}), and between $R=0\rightarrow R_0$, we have,
\be
\int dR \int d\alpha { R e^{{-{(R -R_p)^2\over \sigma ^2}}}
dR d\alpha\over R_0^2 + R^2 \mp 2 R_0 R \cos \alpha}\approx 13.7 \, {\rm kpc}^2 \,.
\ee
Therefore the total solid angle subtended at the solar position by all superbubbles within the limiting longitudes, is
\be
\Omega_{Tot}={13.7 \pi \, \Sigma_0 \, \ln [{r_{s,max} \over r_{s,min}}]  \over \Bigl [r_{s,min}^{-2} -  r_{s,max}^{-2} \Bigr ]} \,.
\ee
The total solid angle within Galactic longitudes ($80^\circ,-80^\circ$) and latitudes $(5^\circ,-5^\circ)$, is $0.9\pi \times 2[1- \cos (5^\circ)]=0.007 \pi$ steradian. Therefore the sky filling factor of all the superbubbles in the Galactic inner region, is,
\be
f_{\rm sky}={2\times 10^{3} \Sigma_0 \ln [{r_{s,max} \over r_{s,min}}]  \over \Bigl [r_{s,min}^{-2} - r_{s,max}^{-2} \Bigr ]} \,
\ee
Taking $r_{s,min}=10$ pc, $r_{s,max}=100$ pc, 
%and the logarithmic factor as $2.3$, for a range of superbubbles sizes $10\hbox{--}100$ pc,
we have,
\be
f_{\rm sky}=0.46 \times \Sigma_0 ({\rm kpc}^{-2})\approx 9 \,.
\ee
Even if the number of star clusters is reduced by an order of magnitude, the sky filling factor would be close to unity. Therefore all lines of sight towards
the Galactic inner region is bound to intercept the superbubble triggered by a young ($\le 10$ Myr) star cluster.

\section{Diffuse gamma-ray emission}
The scaling  of $r_s \propto L_w^{1/2}$ helps us to estimate the diffuse gamma-ray flux from superbubbles, because this implies that the specific intensity of $\gamma$-radiation is same for all superbubbles. Since all lines of sight intercepts at least one superbubble, all we have to do is to calculate
the specific intensity in one superbubbles, and this will suffice to estimate the diffuse gamma-ray flux. The specific intensity is given by 
$\pi \times I_\nu =L_\nu /(4 \pi r_s^2)$, where $L_\nu$ is the  luminosity of a superbubble of radius $r_s$, and the mean specific intensity in the case of isotropic radiation is $J_\nu \equiv I_\nu$. 

The fraction of mechanical wind power of massive stars in clusters that is radiated in GeV $\gamma$-rays varies a lot. 
%{NewCut}
Observations of NGC 3603 and Westerlund 2 suggest a fraction $2 \times 10^{-3}$ and $6 \times 10^{-3}$, while the fraction in 30 Doradus is large, of order $0.01$, and the fraction in Cygnus is small, $\approx 3 \times 10^{-4}$ (see Table 1 of \citet{gupta2018}). 
It can be shown that for a self-similarly evolving superbubble, the fraction depends on the density profile, dynamical time and the fraction of shock energy that is deposited in cosmic rays (equation 3 of \citet{gupta2018}). While for a uniform ambient density, the fraction
scales as $\rho t_{dyn}$, for a density profile of the type $\rho=\rho_c  (r/r_c)^{-1}$ beyond a core radius $r_c$, the fraction scales as $L_w^{1/4} (\rho_cr_c) ^{5/4} t_{dyn}^{1/4}$. Variation of ambient density, compactness ($r_c$), the age of a cluster ($t_{dyn}$) can cause the fraction to differ between cases. Moreover, cosmic ray diffusion coefficient and the fraction of shock energy deposited into cosmic rays can cause further variations.

If we use a fraction as low as $\sim 10^{-3}$, then the diffuse gamma-ray flux at $\sim 1$ GeV is given by (using equation \ref{rs-lw}),
\ba
J_\gamma &\approx& {10^{-3} L_w \over 4 \pi^2 r_s^2} \approx 1.2 \times 10^{-8} {\rm erg} \, {\rm cm}^{-2} \, {\rm s}^{-1} \, {\rm sr}^{-1} \nonumber\\
&\approx&  10^{-2} \, {\rm MeV} \, {\rm cm}^{-2} \, {\rm s}^{-1} \, {\rm sr}^{-1}\,.
\ea
The diffuse gamma-ray flux observed by \cite{ackermann2012} (their Figure 15) is $\approx 4 \times 10^{-2}$ MeV cm$^{-2}$ s$^{-1}$ sr$^{-1}$. 

Clearly the variation in the efficiency of production of $\gamma$-rays, which depend on the ambient density, cosmic ray diffusion coefficient and other factors, will
produce anisotropy. In addition to this, there will be anisotropy from clustering of superbubbles, which can be estimated as follows.
The resolution of Fermi-LAT in GeV range is $0.8^\circ$. We can estimate the number of star clusters intercepted in the solid angle of the beam, of $1.9 \times 10^{-4}$ steradian. Since roughly $3000$ clusters produce a total solid angle of $0.19$ steradian, therefore in the beam, one expects
roughly $N\approx 3\hbox{--}4$ clusters. This implies an anisotropy $1/\sqrt{N}\approx 0.5$. 

The anisotropy is expressed in terms of spherical harmonics decomposition. Consider the angular power spectrum of intensity fluctuation,  $\delta I (\theta)=(I(\theta)-\langle I \rangle)/\langle I \rangle$, where $I(\theta)$ is the intensity in direction $\theta$ and $\langle I \rangle$ is the average intensity. The angular power spectrum is calculated by expanding $\delta I$ in terms of spherical harmonics $\delta I=\Sigma_{l,m} a_{l,m} Y_{l.m} (\theta)$, and the coefficients $C_l=\langle \vert a_{l,m} \vert ^2 \rangle$ then become a measure of anisotropy. For Poisson distribution, $C_l \equiv C^P$, independent of the multipole $l$, and equal to the number of sources intercepted in the instrumental beam. The anisotropy of the diffuse flux observed by {\it Fermi-LAT} has been analysed after masking the Galactic plane ($b< 30^\circ$). For an estimate of the anisotropy at higher latitude, we assume that the above mentioned distribution of star clusters have an exponential profile ($\exp (-z/h)$), with a scale height of $500$ pc, and appropriately normalised. The number density of star clusters at a location $(R,z)$ (cylindrical coordinates), is then given by,
\be 
n\approx {1 \over \Sigma_0 /2 h} \,e^{-{(R -R_p)^2\over \sigma ^2}} \, e^{-z/h} \,.
\ee
Then the number of superbubbles intercepted in a beam at %a given latitude can be estimated. At 
$b=60^\circ$ %, the number of sources is
can be estimated as $\approx 9$ sr$^{-1}$, for a beam %solid angle 
extending up to the Galactic Centre. %(beyond which the additional sources are unlikely to be large). 
Therefore, the number of sources in the {\it Fermi-LAT} beam is $\mathcal{N}\sim 1.7 \times 10^{-3}$, and unlikely to contribute towards the diffuse background at high latitude. It is interesting to note that other Galactic sources are also unlikely to do the same \citep{biswas2019}. 
%The anisotropy arising from star clusters
%would be $C^P \approx 9$ sr, compared to the observed value of $C^P \approx 10^{-5}$ sr \citep{ackermann2012b} \red{ although at those latitudes star clusters are unlikely to be main source of DGE.}

%A further indication that star clusters may be responsible to a substantial fraction of DGE, comes from the analysis by \citet{deboer2017}, who
%found that molecular clouds can explain the DGE flux towards the central molecular zone, with a harder spectrum than that found elsewhere. Interestingly, the observations of \citet{Aharonian2019} found that $\gamma$-ray spectrum towards star clusters have a harder spectrum.

\section{Conclusions}
We have considered the contribution of massive young star clusters to the diffuse Galactic $\gamma$-ray emission. We have shown that the sky filling factor of star cluster triggered superbubbles  towards the inner Galaxy is more than unity. Since their $\gamma$-ray luminosity scales as square of the bubble size, we showed that the resulting surface brightness is comparable to the observed intensity of DGE. %We have also discussed the anisotropy from Possion statistics of unresolved sources and other implications of this scenario. 
Our estimates show that $\gamma$-rays from star clusters can fill the gap between the observed DGE and that predicted from leaky-box model, thereby supporting a de-coupling of grammage with CR propagation in ISM. This model predicts a rather large anisotropy of DGE in the inner
Galactic region which could be tested in future. 

\bigskip
BBN wishes to thank S. Thoudam and P. L. Biermann for useful discussions. We thank an anonymous referee for detailed comments.

\section{Data availability}
The data underlying this article are available in the article.


\begin{thebibliography}{}  
\bibitem[\protect\citeauthoryear{Ackermann et al.}{2011}]{Ackermann2011}
Ackermann, M. \etal  2011
\ufhref[webgreen]{https://doi.org/10.1126/science.1210311}{Science}, 
\ufhref[webgreen]{https://ui.adsabs.harvard.edu/abs/2011Sci...334.1103A}{334, 1103} 

\bibitem[\protect\citeauthoryear{Ackermann et al.}{2012}]{ackermann2012}
Ackermann, M. \etal  2012
\ufhref[webgreen]{https://doi.org/10.1088/0004-637X/750/1/3}{ApJ}, 
\ufhref[webgreen]{http://adsabs.harvard.edu/abs/2011AJ....141...23B}{750, 3} 

\bibitem[\protect\citeauthoryear{Ackermann et al.}{2012b}]{ackermann2012b}
Ackermann, M. \etal  2012b
\ufhref[webgreen]{https://doi.org/10.1103/PhysRevD.85.083007}{PRD}, 
\ufhref[webgreen]{https://ui.adsabs.harvard.edu/abs/2012PhRvD..85h3007A}{85, 3007} 

\bibitem[\protect\citeauthoryear{Aharonian et al.}{2019}]{Aharonian2019}
Aharonian, F., Yang, R., Wilhelmi, E. O.  2019
\ufhref[webgreen]{https://doi.org/10.1038/s41550-0190-0724-0}{Nature}, 


\bibitem[\protect\citeauthoryear{Bagetakos et al.}{2011}]{bagetakos2011}
Bagetakos, I., Brinks, E., Walter, F., De Block, W. J. G., Usero, A., Leroy, A. K., Rich, J. W., Kennicutt, Jr, R. C. 2011
\ufhref[webgreen]{https://doi.org/10.1088/0004-6256/141/1/23}{AJ}, 
\ufhref[webgreen]{http://adsabs.harvard.edu/abs/2011AJ....141...23B}{141, 23} 

\bibitem[\protect\citeauthoryear{Berezhko}{2000}]{berezhko2000}
Berezhko, E. G., V\"olk, H. J.  2000
\ufhref[webgreen]{https://doi.org/10.1086/309354}{A\&A}, 
\ufhref[webgreen]{https://ui.adsabs.harvard.edu/abs/2000ApJ...540..923B}{540, 923} 

\bibitem[\protect\citeauthoryear{Biswas \& Gupta}{2019}]{biswas2019}
Biswas, S., Gupta, N.  2019
\ufhref[webgreen]{https://ui.adsabs.harvard.edu/abs/2019arXiv190703102B}{arXiv:1907.03102}, 

\bibitem[\protect\citeauthoryear{Bloemen}{1989}]{bloemen1989}
Bloemen, H. 1989
\ufhref[webgreen]{https://doi.org/ 10.1146/annurev.aa.27.090189.002345}{ARAA}, 
\ufhref[webgreen]{https://ui.adsabs.harvard.edu/abs/1989ARA\%26A..27..469B}{27, 469} 

\bibitem[\protect\citeauthoryear{Bronfman et al.}{2000}]{Bronfman2000}
Bronfman, L., Casassus, S., May, J., Nyman, L.-\AA . 2000, A\&A
%\ufhref[webgreen]{https://doi.org/10.1088/0004-6256/141/1/23}{A\&A}, 
\ufhref[webgreen]{http://adsabs.harvard.edu/abs/2000A\%26A...358..521B/}{358, 521} 

\bibitem[\protect\citeauthoryear{Chomiuk \& Povich}{2011}]{chomiuk2011}
Chomiuk, L., Povich, M. S. 2011
\ufhref[webgreen]{https://doi.org/10.1088/0004-6256/142/6/197}{AJ}, 
\ufhref[webgreen]{http://adsabs.harvard.edu/cgi-bin/bib_query?2011AJ....142..197C}{142, 197} 



\bibitem[\protect\citeauthoryear{Cowsik \& Madziwa-Nussinov}{2016}]{cowsik2016}
Cowsik, R., Madziwa-Nussinov, T. 2016
\ufhref[webgreen]{https://doi.org/10.3847/0004-637X/827/2/119}{ApJ}, 
\ufhref[webgreen]{https://ui.adsabs.harvard.edu/abs/2016ApJ...827..119C}{827, 119} 

%\bibitem[\protect\citeauthoryear{Clarke \& Oey}{2002}]{clarke2002}
%Clarke, C., Oey, M. S. 2002, MNRAS, 337, 1299
% \ufhref[webgreen]{https://doi.org/10.1093/mnras/sty1846}{MNRAS}, 
 %\ufhref[webgreen]{http://adsabs.harvard.edu/abs/2014ApJ...794..144M}{479, 5220} 
 
 \bibitem[\protect\citeauthoryear{de Boer \etal}{2017}]{deboer2017}
de Boer, W., Bosse, L., Gebauer, I., Neumann, A., Biermann, P. L.  2017
\ufhref[webgreen]{https://doi.org/10.1103/PhysRevD.96.043012}{PRL}, 
\ufhref[webgreen]{https://ui.adsabs.harvard.edu/abs/2017PhRvD..96d3012D}{96, 043012} 

 
 \bibitem[\protect\citeauthoryear{Diehl \etal}{2006}]{Diehl2006}
Diehl, R. \etal 2006
\ufhref[webgreen]{https://doi.org/10.1038/nature04364}{Nature}, 
\ufhref[webgreen]{https://ui.adsabs.harvard.edu/abs/2006Natur.439...45D}{439, 45} 

 \bibitem[\protect\citeauthoryear{Eichler}{2017}]{eichler2017}
Eichler, D. 2017
\ufhref[webgreen]{https://doi.org/10.3847/1538-4357/aa6a11}{ApJ}, 
\ufhref[webgreen]{https://ui.adsabs.harvard.edu/abs/2017ApJ...842...50E}{842, 50} 

 \bibitem[\protect\citeauthoryear{Gabici \etal}{2019}]{gabici2019}
Gabici, S., Evoli, C., Gaggero, D., Lipari, P., Mertsch, P., Orland, E., Strong, A., Vittino, A. 2019
\ufhref[webgreen]{https://doi.org/10.1142/S0218271819300222}{IJMPD}, 
\ufhref[webgreen]{https://ui.adsabs.harvard.edu/abs/2019IJMPD..2830022G}{28, 1930022-339} 
 
 
 \bibitem[\protect\citeauthoryear{Gupta \etal}{2018}]{gupta2018}
Gupta, S., Nath, B. B., Sharma, P. 2018
\ufhref[webgreen]{https://doi.org/10.1093/mnras/sty1846}{MNRAS}, 
\ufhref[webgreen]{http://adsabs.harvard.edu/abs/2011AJ....141...23B}{479, 5220} 

 \bibitem[\protect\citeauthoryear{Gupta \etal}{2020}]{gupta2020}
Gupta, S., Nath, B. B., Sharma, P., Eichler, D. 2020
\ufhref[webgreen]{https://doi.org/10.1093/mnras/staa286}{MNRAS}, 
\ufhref[webgreen]{https://ui.adsabs.harvard.edu/abs/2020MNRAS.493.3159G}{493, 3159} 

 \bibitem[\protect\citeauthoryear{Hunter \etal}{1997}]{hunter1997}
Hunter, S. D \etal 1997
\ufhref[webgreen]{https://doi.org/10.1086/304012}{ApJ}, 
\ufhref[webgreen]{https://ui.adsabs.harvard.edu/abs/1997ApJ...481..205H}{481, 205} 

 \bibitem[\protect\citeauthoryear{Hunter \etal}{1997}]{hunter1997b}
Hunter, S. D, Kinzer, R. L., Strong, A. W. 1997
\ufhref[webgreen]{https://doi.org/10.1063/1.54029}{The fourth Compton symposium. AIP Conference Proceedings}, 
\ufhref[webgreen]{https://ui.adsabs.harvard.edu/abs/1997AIPC..410..192H}{410, 192} 

\bibitem[Jenkins \& Tripp (2011)]{jenkins2011}
Jenkins, E. B. \& Tripp, T. M.  2011,
 \ufhref[webgreen]{http://dx.doi.org/10.1088/0004-637X/734/1/65}{ApJ}, 
 \ufhref[webgreen]{http://adsabs.harvard.edu/abs/2011ApJ...734...65J}{734, 65}

\bibitem[Seo \etal (2018)]{kang2018}
Seo, J., Kang, H., Ryu, D.  2018,
 \ufhref[webgreen]{http://dx.doi.org/10.5303/JKAS.2018.51.2.37}{JKAS}, 
 \ufhref[webgreen]{https://ui.adsabs.harvard.edu/abs/2018JKAS...51...37S}{51, 37}


\bibitem[\protect\citeauthoryear{Krause \& Diehl}{2014}]{krause2014}
 Krause, M. G. H., Diehl, R. 2014
 \ufhref[webgreen]{https://doi.org/10.1088/2041-8205/794/2/L21}{ApJ}, 
 \ufhref[webgreen]{http://adsabs.harvard.edu/abs/2014ApJ...794..144M}{794, L21} 
 
 
\bibitem[\protect\citeauthoryear{Lopez et al}{2014}]{lopez2014}
Lopez, L. A., Krumholz, M. R., Bolatto, A. D., Prochaska, J. X.,  Ramirez-Ruiz, E. \& Castro, D. 2014,
 \ufhref[webgreen]{http://dx.doi.org/10.1088/0004-637X/795/2/121}{ApJ},
 \ufhref[webgreen]{http://adsabs.harvard.edu/abs/2014ApJ...795..121L}{795,121}

\bibitem[\protect\citeauthoryear{McKee \& Williams}{1997}]{mckee1997}
 McKee, C. F., Williams, J. P. 1997
 \ufhref[webgreen]{https://doi.org/10.1086/303587}{ApJ}, 
 \ufhref[webgreen]{http://adsabs.harvard.edu/abs/1997ApJ...476..144M}{476, 144}  
 
 \bibitem[\protect\citeauthoryear{Morrison}{1958}]{morrison1958}
Morrison, P. 1958
 \ufhref[webgreen]{https://doi.org/10.1093/mnras/staa336}{Nuovo Cimento}, 
 \ufhref[webgreen]{https://ui.adsabs.harvard.edu/abs/2020MNRAS.493.1034N}{7, 858} 
 
  \bibitem[\protect\citeauthoryear{Nath}{2020}]{nath2020}
Nath, B. B., Das, P., Oey, M. S. 2020
 \ufhref[webgreen]{https://doi.org/10.1007/BF02745590}{MNRAS}, 
 \ufhref[webgreen]{https://ui.adsabs.harvard.edu/abs/1958NCim....7..858M/abstract}{493, 1034} 
 
 
 \bibitem[\protect\citeauthoryear{Oey \& Clarke}{1997}]{oey1997} 
Oey, M. S., Clarke, C. J. 1997
\ufhref[webgreen]{https://doi.org/10.1093/mnras/289.3.570}{MNRAS}, 
 \ufhref[webgreen]{http://adsabs.harvard.edu/abs/1997MNRAS.289..570O}{289, 570} 
 
 
  \bibitem[\protect\citeauthoryear{Reed \etal}{2005}]{Reed2005} 
Reed, B. C. 2005
\ufhref[webgreen]{https://doi.org/ 10.1086/444474}{AJ}, 
 \ufhref[webgreen]{https://ui.adsabs.harvard.edu/abs/2005AJ....130.1652R}{289, 570} 
 
  \bibitem[Sharma \etal(2014)]{sharma2014}
Sharma, P., Roy, A., Nath, B. B., Shchekinov, Y. 2014,
 \ufhref[webgreen]{http://dx.doi.org/10.1093/mnras/stu1307}{MNRAS}, 
 \ufhref[webgreen]{http://adsabs.harvard.edu/abs/2014MNRAS.443.3463S}{443, 3463}
 
 \bibitem[\protect\citeauthoryear{Strong \etal} {2010}]{strong2010}
Strong, A. W., Porter, T. A., Digel, S. W., J\'ohannesson, G., Martin, P., Moskalenko, I. V., Murphy, E. J., Orlando, E. 2010
 \ufhref[webgreen]{http://dx.doi.org/10.1088/2041-8205/722/1/L58}{ApJL},
 \ufhref[webgreen]{https://ui.adsabs.harvard.edu/abs/2010ApJ...722L..58S}{722, L58}
 
   \bibitem[Seo \etal(2018)]{seo2018}
Seo J., Kang, H., Ryu, D. 2018,
 \ufhref[webgreen]{http://dx.doi.org/10.5303/JKAS.2018.51.2.37}{JKAS}, 
 \ufhref[webgreen]{https://ui.adsabs.harvard.edu/abs/2018JKAS...51...37S}{51, 37}
 
 \bibitem[\protect\citeauthoryear{Weaver \etal} {1977}]{weaver1977}
Weaver, R., McCray, R., Castor, J., Shapiro, P., Moore, R., 1977,
 \ufhref[webgreen]{http://dx.doi.org/10.1086/155692}{ApJ},
 \ufhref[webgreen]{http://adsabs.harvard.edu/abs/1977ApJ...218..377W}{218, 377}
 
  \bibitem[\protect\citeauthoryear{Weekes \etal} {1997}]{weekes1997}
Weekes, T. C. \etal 1997
 \ufhref[webgreen]{http://dx.doi.org/10.1063/1.54035}{The fourth Compton symposium. AIP Conference Proceedings},
 \ufhref[webgreen]{http://https://ui.adsabs.harvard.edu/abs/1997AIPC..410..361W}{410, 361}


\end{thebibliography}
\end{document}